\documentstyle[12pt,aasms4]{article}
\received{11 June 1997}
\accepted{6 August 1997}
\slugcomment{
To be appeared in The Astronomical Journal, Vol. 114, No. 5 (1997 Nov issue)
}

\begin{document}

\newcommand {\degr}{$^{\circ}$}
\newcommand {\etal}{{\it et al.} }
\newcommand {\UTT}{$U_{\rm T}$}
\newcommand {\BTT}{$B_{\rm T}$}
\newcommand {\VTT}{$V_{\rm T}$}
\newcommand {\UBVTT}{$(U-B-V)_{\rm T}$}
\newcommand {\BTo}{$B_{\rm T}^{0}$}
\newcommand {\UBVTo}{$(U-B-V)_{\rm T}^{0}$}
\newcommand {\UBTT}{$(U-B)_{\rm T}$}
\newcommand {\BVTT}{$(B-V)_{\rm T}$}
\newcommand {\UVTT}{$(U-V)_{\rm T}$}
\newcommand {\UBTo}{$(U-B)_{\rm T}^{0}$}
\newcommand {\BVTo}{$(B-V)_{\rm T}^{0}$}
\newcommand {\UVTo}{$(U-V)_{\rm T}^{0}$}
\newcommand {\fFIRfB}{$f_{FIR}/f_{B}$}
\newcommand {\ftenfsix}{$f_{100}/f_{60}$}
\newcommand {\HI}{H~{\small I}}
\newcommand {\HII}{H~{\small II}}
\newcommand {\VGSR}{$V_{\rm GSR}$}
\newcommand {\LB}{$L_{B}$}
\newcommand {\Lsolar}{$L_{\odot}$}
\newcommand {\Msolar}{$M_{\odot}$}
\newcommand {\HB}{H$\beta$}
\newcommand {\HD}{H$\delta$}
\newcommand {\fFIR}{$f_{FIR}$}
\newcommand {\LFIR}{$L_{FIR}$}
\newcommand {\MT}{$M_{T}$}
\newcommand {\MTLB}{$M_{T}/L_{B}$}
\newcommand {\LFIRMT}{$L_{FIR}/M_{T}$}
\newcommand {\LFIRLB}{$L_{FIR}/L_{B}$}

\title{
Characteristics of Kiso Ultra-Violet Excess Galaxies
}

\author{
Akihiko Tomita
}
\affil{
Department of Earth and Astronomical Sciences,
\\
Faculty of Education, Wakayama University,
\\
930 Sakae-Dani, Wakayama 640, Japan
\\
Electronic mail: atomita@center.wakayama-u.ac.jp
}

\and

\author{
Tsutomu T. Takeuchi\altaffilmark{1},
Tadashi Usui, and
Mamoru Sait\={o}
}
\affil{
Department of Astronomy,
Faculty of Science, 
Kyoto University,
\\
Sakyo-ku, Kyoto 606-01, Japan
\\
Electronic mail:
takeuchi, usui, saitom@kusastro.kyoto-u.ac.jp
}

\altaffiltext{1}{
Research Fellow of the Japan Society for the Promotion of Science
} 

\begin{abstract}
We examined the general characteristics
of the Kiso Ultra-violet Excess Galaxies (KUGs).
We present for the first time the quantitative expressions
for the criteria of the KUGs;
the boundary color separating the KUGs from the non-KUGs
is \BVTT\ =~0.74
%  and the mean \BVTT\ colors for objects with the UV degrees
%  of H, M, and L are 0.48, 0.54, and 0.67, respectively.
and the KUG degrees of UV strength are found to correlate
with the mean \BVTT\ colors.
We investigate the nature of the KUGs,
a sample of blue galaxy population,
and show that
(1) about a half of the KUGs are spiral galaxies
with Sb to Scd,
(2) the KUGs are biased to late-type galaxies and
include early-type galaxies with young star populations, and
(3) the KUGs are preferably found among less luminous galaxies
with \LB\ $<$~$10^{10}$\Lsolar.
The KUGs
also contain the post-starburst galaxies,
many of which are found among the blue galaxy population
at intermediate redshifts.
The analysis of the far-infrared data shows that
a typical present-to-past star formation rate for a KUG is 0.4.
\end{abstract}


% Section 1
\section{Introduction}

Kiso Ultraviolet-Excess Galaxies (hereafter KUGs) are
a UV-excess galaxy sample surveyed 
by means of the plates taken by the Kiso Observatory
1.05-m Schmidt telescope,
the compiled catalog of which was given by
Takase, Miyauchi-Isobe (1993) (TM93).
KUGs were selected by eye-inspection among galaxies
on the Schmidt plates made by the $U, G, R$ three-image method
if the objects had $U$-images brighter than $G$ and $R$-images;
the exposure times were so adjusted that
an A0 star would have an equal brightness
for the three images.
The survey method,
however,
did not introduce quantitative criteria of the KUGs.

Many follow-up observations were made
to investigate the characteristics of the KUG.
These are a radio continuum study of 38 selected objects
(Maehara \etal 1985),
an optical spectroscopic study of 57 selected objects
(Maehara \etal 1987),
optical and \HI\ observations of four selected objects
(Maehara \etal 1988),
and optical and \HI\ observations of 142 compact-morphology
objects (Augarde \etal 1994; Comte \etal 1994).
The results revealed that
the KUGs are similar to the Markarian galaxies
(Mazzarella, Balzano 1986)
and most of the KUGs are actively star-forming galaxies.
However,
these previous studies did not examine the whole KUG sample;
the samples used were much smaller in number
compared with the total number of the KUG of 8162,
or taking only a subsample with a certain morphology.

In this paper,
using known optical color data,
morphological data,
and far-infrared data of a large number of the KUGs,
we clarify the general characteristics of the KUGs quantitatively.
We describe the sample and data in section~2.
The optical colors and the morphological characteristics
in connection with the Hubble system are given in section~3.
In section~4,
we discuss the nature of the KUGs
by analyzing the optical luminosity functions,
the far-infrared data,
and some KUGs in Coma cluster.
A summary is given in section~5.

% Section 2
\section{Data Description}

The KUG survey was carried out in 170 Schmidt fields so far.
Each field has a view of 5.7\degr $\times$ 5.7\degr,
about 32 deg$^{2}$,
and the total area of the survey is about 5100 deg$^{2}$,
or 1.55 steradian.
The total number of KUGs is 8162.
A map of the surveyed regions was presented
in figure~1 of TM93.
TM93 noted that
the KUG survey is not homogeneous in depth
from the survey plate to plate;
the limiting magnitude varies from $m_{pg}$ = 16.5 to 18 mag
depending on the plate quality.
In order to avoid the effect of the inhomogeneity,
we restrict a KUG sample used in this paper
to those contained in the Zwicky catalog
(Zwicky \etal 1961 -- 1968 [CGCG]; hereafter CGCG galaxies),
because the limiting magnitude of the CGCG galaxies
is $m_{mg}$ = 15.7,
much less deep than that of the KUGs;
hereafter we simply call them KUGs.
We make a reference sample
which are CGCG galaxies not selected as KUGs in the KUG-survey regions;
we call them non-KUGs.
Among all 170 KUG-survey fields,
there are eight southern fields out of the CGCG-survey region
of $\delta > -5$\degr.
The A0432 region (center position: $16^{\rm h}20^{\rm m}$, + 35\degr)
is excluded in the following analyses
because of the failure of the survey
(Miyauchi-Isobe, Takase, Maehara 1997).
Thus we take 161 fields out of 170.
There are 2374 KUGs and 5797 non-KUGs in these regions.

We take the data for optical colors
and Hubble-system morphological types,
$T$ index,
from the RC3 (de Vaucouleurs \etal 1991).
Therefore,
our sample is restricted to those which are cataloged in the RC3.
From the KUG catalog,
we take the KUG morphology and the UV degree,
which are given for all KUGs.
We use the $IRAS$ data,
therefore,
we exclude objects located at the $IRAS$ non-survey regions
and close binary galaxies to each of which
the $IRAS$ measurements can not assign the far-infrared fluxes.
The numbers of KUGs and non-KUGs in our sample become 1246 and 2804,
respectively.
Note that
all of the objects do not always have full listings in data.

% Section 3
\section{Optical Characteristics of the KUGs}

%% Section 3.1
\subsection{Color Difference between the KUGs and the non-KUGs}

Since the picking up of the KUGs was made
referring to colors of A0 stars on the same plate,
the selection of the KUGs was little affected
by the Galactic extinction;
in analyzing the color data,
we use mainly the color systems without the extinction correction.
Figure~1 shows the color histograms for the KUG (solid lines)
and the non-KUG (dashed lines) samples
in \UBVTT\ system.
Each histogram is normalized to the total number of each sample.
The color distributions differ between the KUGs and the non-KUGs.
The number of objects,
mean color,
standard deviation of the distribution,
and the colors at 20\% and 80\%-level of frequency distribution
in each color system are summarized in table~1.
We indicate an arrow in each histogram as the boundary between
KUG and non-KUG colors,
at which the fraction of KUGs with redder colors
is equal to that of non-KUGs with bluer colors.
We call the fraction of KUGs with the non-KUG colors,
or the fraction of non-KUGs with the KUG colors,
the overlapping fraction,
which are 21\% to 23\% for all the colors.
The boundary colors are 0.10 in \UBTT\
and 0.74 in \BVTT.
The color at the boundary and the overlapping fraction
in each color are shown in the last two columns of table~1.
These are the first quantitative expressions
for the color criterion for the KUG.

Though the overlapping fractions are less than one fourth,
the colors of the overlapped regions in the KUG sample
extend up to the peak colors in the non-KUG sample.
This may be due to the property of the survey procedure,
i.e., eye-inspection of the three-image plates.
For instance,
the KUG survey might pick up objects with blue knots
which are not always blue in total color.

Figure~2 shows $U-B$ versus $B-V$ color-color diagram
for the KUGs and the non-KUGs.
Figure~2a shows in \UBVTT\ system.
Circles indicate the KUGs of which both \UBTT\ and \BVTT\ colors
are given in the RC3,
216 objects out of total 1246,
and crosses indicate the non-KUGs,
427 objects out of total 2804.
The boundary between KUG and non-KUG regions is shown as a dashed line,
which has been defined in figure~1 to be the boundary color
of \UVTT\ = 0.83.
Figure~2b shows the color variation of the sample galaxies
of both the KUGs and the non-KUGs
along the morphology sequence in the Hubble system.
The positions of dots indicate the mean colors
and the error bars indicate
the standard deviations of the color distributions.
The color data and binning of the morphological types are summarized
in table~2.
From figures~2a and 2b,
we find that
the colors of the KUGs correspond
to those of Sc or later-type galaxies.

Figures~2c is the same as figure~2b,
but for the \UBVTo\ system.
The numbers of plotted galaxies,
having both \UBTo\ and \BVTo\ colors in the RC3,
are 208 and 403 for KUGs and non-KUGs,
respectively.
The boundary between KUG and non-KUG regions
indicates the relation of \UVTo\ = 0.64,
which is derived by the same procedure as made in \UVTT\ system.
The colors of the main sequence stars are plotted in figure~2c
following Allen (1973).
Though the KUG colors were originally intended
to be bluer than or equal to those of A0 stars,
the color boundary has actually corresponded to G0 stars.

The KUG researchers introduced the UV degree,
i.e., blueness,
of the KUG images into three classes,
H, M, and L from high to low.
Figure~3 shows the color distribution of the sample
in each UV degree in two color systems,
\UBTT\ and \BVTT.
The number of sample,
mean color,
and standard deviation of the color distribution
is summarized in table~3.
The mean \BVTT\ colors for the H, M, and L samples are
0.48, 0.54, and 0.67, respectively,
showing a correlation between the UV degree and the total colors.
However,
the correlation is not tight as shown in figure~3;
this may be due to the property of the KUG survey procedure.
Though the colors of the KUGs with the UV degree of L
overlaps those of the non-KUGs,
the mean colors of L are still bluer than
the boundary colors separating the KUGs from the non-KUGs;
for instance,
the mean \BVTT\ color for objects with L is 0.67,
while the boundary color is 0.74.

%% Section 3.2
\subsection{KUG Fraction as a Funcion of the Hubble Sequence}

The fraction of the KUG for each $T$ index
along the Hubble sequence is tabulated in table~4.
The binning of the $T$ index is shown in the first column.
Another notation for the morphology is shown in the second column;
$-5$: E, $-2$: S0, 1: Sa, 3: Sb, 5: Sc, 7: Sd, 9: Sm,
10: Im, 11: cI (no object in our sample), 90: I0, and 99: peculiar.
Figure~4 illustrates the data in table~4.
The abscissa indicates the $T$ index of morphology given in the RC3;
$-$6 to 11, one space, 90 and 99, and one space.
The solid histogram indicates the KUGs in number
and the dashed histogram indicates the total sample in number.
The number scale is shown in the left-side ordinate.
The solid broken line indicates the KUG fraction along the Hubble sequence.
The scale for the fraction in \% is shown in the right-side ordinate.
In number,
54\% of the KUGs with $-6 \leq T < 11$
have the Hubble types with
3~$\leq$~$T$~$<$~7;
a majority of the KUGs are spirals of Sb or Sc.

The fraction is monotonously increasing from early to late types.
For E/S0 galaxies ($T$ $<$ 0) the fraction is less than 10\%,
on the other hand,
for Sd -- Im galaxies ($T$ = 7 -- 10)
the fraction is about 50\%.
For $T \geq$ 5 (later than or equal to Sc),
the fraction exceeds 40\% and this trend is consistent with the results
by the color analysis given in section~3.1,
i.e.,
the colors of the KUGs are bluer than or equal to those of Sc.
However,
we should note the following two points.
Though the number is small,
some E/S0 galaxies are recognized as KUGs.
Though the fraction is high,
about a half of Sd -- Im galaxies are not recognized as KUGs.

Figure~5 shows the frequency of the UV degree along the Hubble sequence,
and the data for figure~5 is also tabulated in table~4.
The fraction of objects with higher UV degrees is monotonously
increasing with the $T$ index.
It is found from figures~2, 4, and 5 that
the KUGs are biased to the late-type galaxies in the Hubble sequence,
and another contribution to the KUGs are 
galaxies with young star population regardless of the Hubble type.

Picking up of the KUG may be affected by the galaxy morphology
as is mentioned in section~3.1;
objects with blue knots may tend to be picked up,
even if they are red in total color.
The \BVTT\ color distribution
for objects with the UV degree of M
is shown in figure~6 and tabulated in table~5
along the Hubble sequence.
A color variation is found in figure~6;
the later the Hubble type is,
the bluer the color is.
The earliest category,
$-6 \leq T < 2$,
has a mean color of 0.66,
while the latest category,
$7 \leq T \leq 11$,
has a mean color of 0.46,
which is 0.2~mag bluer than the former one.
This means that the early-type KUGs tend to show
the peculiar features of having blue knots in red disks or bulges.

The color difference in \BVTT\ of 0.2 mag
between the earliest and the latest KUGs is smaller than
that of 0.5 mag
in the non-KUGs as shown in the fifth column of table~5.
Table~5 shows that
the color difference between the KUG and the non-KUG
becomes larger in earlier-type galaxies;
the color difference is larger than 0.2~mag
for galaxies with $T$ $<$ 5,
and it reduces to less than 0.04~mag for galaxies with $T$ $\geq$ 5.
This suggests that
the late-type KUGs consist of mostly normal galaxies
for their morphologies
in terms of the stellar population,
while the early-type KUGs consist of galaxies
containing much young star populations for their morphologies.

%% Section 3.3
\subsection{Morphological Characteristics of the KUGs}

The KUG catalog provides an original morphological classification
(Takase, Noguchi, Maehara 1983) as follows;
Ic is an irregular galaxy with clumpy \HII\ regions,
Ig is an irregular galaxy with a conspicuously giant \HII\ region,
Pi is a pair of interacting components,
Pd is a pair of detached components,
Sk is a spiral galaxy with knots of \HII\ regions along arms,
Sp is a spiral galaxy with peculiar bar and/or nucleus,
C is a compact galaxy,
and `?' is an unclassifiable one.
TM93 summarized the frequency
of the KUG morphological system and that of the UV degree in each
KUG morphological type.
In all following morphological analyses,
we include objects with suspect classification;
for instance,
subsample with Ic is a combined sample of objects assigned
originally as `Ic' and `Ic:'.

The correlation between the KUG morphology and the Hubble system,
$T$ index,
is tabulated in table~6.
It is found from table~6 that
spiral KUGs with 0 $\leq$~$T$~$<$~9 (S0a to Sdm)
are mostly classified as Sk or Sp,
i.e., they have knots of the \HII\ regions in spiral arms or
have peculiar bars or nuclei,
showing that the selection of spiral KUGs were
performed to be matched to these morphologies.
This makes the KUG fraction suppressed to about a half
even in late-type spirals
by excluding the spirals with smoother features in blue light,
though there is no difference in the stellar population
between the late-type KUGs and the late-type non-KUGs
as mentioned in section~3.2.
In the case of later KUG with 9~$\leq$~$T$~$<$~11 (Sm and Im),
contribution of irregular or interacting features is larger.
The earlier KUG with $T$~$<$~0,
the fraction of objects with C and Sp is larger,
indicating the peculiar features,
which is consistent with the indication that
the early-type KUGs tend to have blue knots in red disks or bulges
as mentioned in section~3.2.

It is found from table~6 that most of objects with C
are unclassified in the Hubble system
and about a half of galaxies in the C sample
with known Hubble types are E/S0 galaxies.
Comte \etal (1994) argued that
the C sample consists of distant early-type galaxies.
The radial velocity data of \VGSR\ in the RC3,
a velocity system measured in the Galactic Standard of Rest,
is given for 104 objects with C out of total 107.
The mean radial velocity for the C sample is \VGSR\ = 5928~km~s$^{-1}$,
similar to that for the total sample,
\VGSR\ = 5224~km~s$^{-1}$;
the C sample is not distant galaxies among the KUGs.
Augarde \etal (1991) showed that
most of the KUGs have the equivalent widths of the \HB\ emission
smaller than 15~\AA,
which are much smaller than that for I~Zw~18 of 40~\AA.
Comte \etal (1994) showed that
\HB\ equivalent widths of the KUGs are systematically smaller than
those of the blue galaxy sample
given by the group at the University of Michigan,
and they claimed that the KUGs are not strong star-forming galaxies.
However,
we should note that their sample
is biased on KUGs with morphology of C.
There are KUGs without the KUG morphology,
though the sample number is small compared with to total number
(see table~6).
The distribution of their $T$ index is flat from E to Im
avoiding the mid-type spirals, Sb to Sc;
this is the inverse trend of the distribution
for the total sample.
The mean radial velocity of the unclassifiable objects
is \VGSR\ = 5274~km~s$^{-1}$,
nearly the same as the mean value for the total sample.
Therefore,
their distances are not too distant to make the classification difficult.

% Section 4
\section{Discussion}

%% Section 4.1
\subsection{Luminosity Function of the KUGs}

We investigate the difference of luminosity functions
between the KUG and the non-KUG samples.
We made the luminosity functions in a way as follows.

We consider $B$-band luminosity \LB\
derived from \BTo.
The RC3 gives both values of \BTo\ and \VGSR\
for 846 galaxies
out of 4050 of the total KUG and non-KUG sample.
Figure~7 shows the distribution of radial velocity
of the sample
in a bin of 1000~km~s$^{-1}$.
The solid curve indicates the canonical $N$-$z$ relation
expected for a uniform distribution of galaxies,
drawn by using the Schechter-type luminosity function
(Schechter 1976) with parameters given by Efstathiou, Ellis, Peterson (1988)
and the limiting magnitude of 14.2 mag,
and normalized by number.
The overabundance at \VGSR\ $<$ 2000~km~s$^{-1}$ is partially due to
the Virgo cluster.

Figure~8 shows the histogram of \BTo\ magnitude
of a combined sample of the KUGs and the non-KUGs
in a bin of 0.2~mag.
The data number used for figure~8 is 3391
out of 4050 of the total sample.
The ordinate is shown in logarithmic scale.
In a range from 11 to 14 mag,
the number distribution is expressed in a single power-law,
log~$N \propto$ dex~(0.5\BTo),
a little shallower than a canonical value of dex~(0.6\BTo)
which is expected for a complete uniform distribution.
The shallower slope may correspond to the overabundance of
near galaxies of \VGSR\ $<$ 2000~km~s$^{-1}$
as shown in figure~7.
The power law is followed down to \BTo\ = 14.2 mag,
therefore,
we take \BTo\ = 14.2 as a limiting magnitude for completeness.

We use a relation between the $B$-band magnitude \BTo\
and $B$-band luminosity \LB,

log~\LB\ [\Lsolar] = $-$0.4 \BTo\ + 2 log~$d$ [Mpc] + 11.968,

\noindent
which was used in Tomita, Tomita, Sait\={o} (1996),
where $d$ is the distance to the galaxy in Mpc.
We define the distance to the galaxy from \VGSR,

$d$ [Mpc] = \VGSR\ [km~s$^{-1}$] / 75.0.

\noindent
If \VGSR\ is less than 75~km~s$^{-1}$,
we assign a distance of 1~Mpc.
The distance where a galaxy with a luminosity of \LB\ is
observed as a 14.2-mag object in \BTo,
$d_{\rm max}$, is expressed as

log~$d_{\rm max}$ [Mpc] = 0.5 log~\LB\ [\Lsolar] $-$ 3.144.

\noindent
We restricted objects of which the distance is less than
$d_{\rm max}$ defined above.
For a given luminosity,
we define a volume,
$(4 \pi d_{\rm max}^{3} / 3) (\Omega / 4 \pi)$,
in which we can surely observe a galaxy with the luminosity;
where $\Omega$ is the solid angle of the effective surveyed area
for our sample galaxies
and we take $\Omega$ =~1.4~sr.
Dividing the number of galaxies in each luminosity bin
by the volume defined above,
we made a luminosity function.

Figure~9 shows the luminosity functions for the KUG sample (solid line)
and the non-KUG sample (dashed line).
The abscissa indicates log~\LB\ [\Lsolar]
and the ordinate indicates the number of galaxies
per Mpc$^{3}$ in a bin of 1 mag.
The solid curve indicates the canonical luminosity function
drawn by using the Schechter function
(Schechter 1976) with parameters given by Efstathiou, Ellis, Peterson (1988).
The luminosity functions of the KUGs and the non-KUGs also
show a Schechter-like behavior
and the luminosity function of a combined sample of the KUGs
and the non-KUGs matches the canonical one well.
For log~\LB\ $<$ 10,
the slope and amplitude of the luminosity function
are the same for both the KUG and non-KUG samples.
The number densities of KUG and non-KUG are
the same in this luminosity range.
On the other hand,
the luminosity function of the KUGs falls faster
than that of the non-KUGs for log~\LB\ $\geq$ 10.
In this luminosity range,
the number density of the KUGs is several times
lower than that of the non-KUGs;
the KUGs are preferably less luminous galaxies
with log~\LB\ $<$ 10.

Figure~10 shows the luminosity functions of the subsamples of the KUGs;
four subsamples divided by the Hubble-type morphology,
$-6 \leq T < 2$ (Sa or earlier),
$2 \leq T < 5$ (Sb or Sbc),
$5 \leq T < 7$ (Sc or Scd),
and $7 \leq T \leq 11$ (Sd or later).
The numbers of the samples are 78, 110, 131, and 94, respectively.
The dashed histogram shows the luminosity function
of the total KUG sample.
Even for the KUGs,
each morphological category shows a specific shape of the luminosity
function as presented by Bingelli, Sandage, Tamman (1988).
The luminosity function for the sample of Sa or earlier
has a steep extension in log \LB\ $<$~9
and this is caused by the population of dwarf ellipticals.
Figure~11 shows the relative fraction of the subsamples
drawn by using the data in figure~10.
For log \LB\ $\geq$~9.4,
a combined sample of Sb to Scd dominates.
The KUGs with luminosities
at around the knee of the KUG luminosity function
are mostly mid-type spirals.
For log \LB\ $\leq$~9.3 the dwarf population,
the sample of Sd or later and the sample of dwarf ellipticals,
dominate.

The apparent KUG fraction would decrease with distance
since the KUG fraction decreases with increasing luminosity
as shown in figure~9.
Figure~12 shows the KUG fraction as a function of \VGSR\
in a bin of 1000~km~s$^{-1}$.
For galaxies with \VGSR\ $<$~4000~km~s$^{-1}$,
the KUG fractions are higher than 40\%,
which is consistent with the fact that the number densities
of the KUGs and the non-KUGs are similar to each other
for fainter galaxies.
The KUG fraction decreases steeply
in a range of \VGSR\ = 4000~km~s$^{-1}$ to 6000~km~s$^{-1}$.
For galaxies with \VGSR\ =~6000 to 13000~km~s$^{-1}$,
the KUG fractions are about one fourth,
which is consistent with that the number density of the KUGs
are several times smaller than that of the non-KUGs
for galaxies brighter than \LB $>$~$10^{10}$\Lsolar.
At \VGSR\ = 6000~km~s$^{-1}$,
the distance is assumed to be 80~Mpc,
a galaxy with log~\LB\ [\Lsolar] =~10 would be seen
as an object with \BTo\ =~14.4,
which is nearly the completeness limit of our sample
(see figure~8).
The variation of the KUG fraction with distance is consistent with
the difference of luminosity functions of the KUGs and the non-KUGs.

%% Section 4.2
\subsection{Far-Infrared Characteristics of the KUGs}

The far-infrared (FIR) emission measured by $IRAS$
has been used as an indicator for the star formation
(e.g., de Jong \etal 1984; Soifer \etal 1987; Bothun, Lonsdale, Rice 1989),
though is has been pointed out that radiation from old stellar populations
should contribute to the total FIR emission in the galaxy
(e.g., Sauvage, Thuan 1992; Walterbos, Greenawalt 1996).
Tomita, Tomita, Sait\={o} (1996) showed that the FIR-to-$B$ band flux ratio,
\fFIRfB,
is a useful measure of the present star formation activity in galaxies.
They studied the variation of the activity in spiral galaxies
analyzing as many as 1681 spirals and found that
the range of log~\fFIRfB\ is from $-$1.5 to 0.5.
Spirals with log~\fFIRfB\ $>$~0 thus represent
the very active subsample.
Following the method used in Tomita, Tomita, Sait\={o} (1996),
we calculated \fFIRfB\ for our sample.
The data of \fFIR\ was
taken from $IRAS$ catalogs,
PSC, FSC, SSSC, and CGQIRAS,
and Rice \etal (1988) and Soifer \etal (1987).
Since we adopted \BTo\ system for the $B$-band magnitude,
only objects with measured \BTo\ in the RC3 are considered here;
the numbers of them are 592 and 1397 for
the KUG and the non-KUG samples,
respectively.
Among them 378 of the KUG sample and 514 of the non-KUG sample
have measured values of \fFIRfB\ and
others have only upper-limit values of \fFIRfB\
because of non-detection of the $IRAS$ measurements.

Figure~13 shows the distribution of log~\fFIRfB\
showing in each Hubble morphological type.
The left-side panels indicate the KUG samples
and the right-side panels indicate the non-KUG samples.
The solid histograms show the samples with measured \fFIRfB\
and the dashed ones show those with only upper-limit values of \fFIRfB.
Note that each histogram is a simple frequency of our sample
and not one for a volume-limit sample as presented
in Tomita, Tomita, Sait\={o} (1996).
With a bias that FIR-bright objects can be seen from more distant places,
the simple histogram may shift to a higher value in \fFIRfB\
compared with one for a volume-limit sample.

It is found from figure~13 that
the KUG samples have several times larger values of \fFIRfB\
than the non-KUG samples for E/S0 and Sa.
On the other hand,
the distributions of the KUG and the non-KUG samples in \fFIRfB\
are similar to each other for Sb or later-type categories.
The fraction of galaxies showing intense activity,
here we take those with log~\fFIRfB\ $\geq$ 0,
is larger in the KUG sample than in the non-KUG sample
for all the morphological types
and especially it is remarkable for E/S0 and Sa samples.

Figure~13 shows a trend that
the median value of \fFIRfB\ is decreasing
from early to late types of the KUG sample.
The flux ratio of \fFIRfB\ is the luminosity ratio of \LFIRLB.
Taking the far-infrared luminosity,
\LFIR,
is a measure of the present star formation rate (SFR)
(Devereux, Young 1991; Sauvage, Thuan 1992),
SFR is expressed as

SFR [\Msolar\ yr$^{-1}$] = $1.4 \times 10^{-10}$ \LFIR\ [\Lsolar].

\noindent
Assuming that the age of the galaxy is $10^{10}$~yr
and that the luminous-to-dark mass ratio is unity
as suggested by Rubin (1987),
the past average SFR,
$<$SFR$>$,
is expressed as \MT\ / $2 \times 10^{10}$.
Then,

SFR / $<$SFR$>$ = 2.8 \LFIRMT.

\noindent
The $B$-band luminosity,
\LB,
is affected by the star formation history
(e.g., Gallagher, Hunter, Tutukov 1984),
though it is nearly a measure of the galaxy mass
(e.g., Hodge 1993).
We make a color-dependent \MTLB\
using \MTLB\ and \BVTo\ data for 245 galaxies
in Nearby Galaxies Catalog (Tully 1988 [NBGC])
with morphologies of $T$ = 1 -- 8.
The extinction correction is made for
the $B$-band luminosity cataloged in the NBGC.
Excluding nine objects by the 3-$\sigma$ rejection,
we got a relation,

log~(\MTLB) [\Msolar\ / \Lsolar] = 0.847 \BVTo\ + 0.188 $\pm$ 0.191.

\noindent
These two equations yield a relation,

log~(SFR / $<$SFR$>$) = log~(\fFIRfB) $-$ 0.847~\BVTo\ + 0.635.

\noindent
The mean values of \BVTo\ for E/S0, Sa, Sb, and Sc in the KUG sample
are 0.73, 0.58, 0.56, and 0.49,
respectively.
Then we found that the median value of log~\LFIRMT\
is around $-$0.9 regardless of the morphology,
and that
the median value of the present-to-past SFR for the KUGs
is derived to be about 0.4.
Kennicutt, Tamblyn, Congdon (1994) studied the present-to-past SFR
which they called b-parameter.
Though we need attention in comparison
between our rough estimation and their calculations
because their analysis was careful
by taking into account the gas recycling process,
our preset-to-past SFR of 0.4 is
larger than their values for Sb or earlier-type galaxies
(see their figure~6).

%% Section 4.3
\subsection{Post-Starburst Galaxies in the KUGs}

The KUGs appear even in early-type galaxies
and these are blue for their Hubble-type morphology
as described in section~3.
We show that post-starburst galaxies
contribute to the early-type KUGs.

Takase (1980) showed the spatial distribution of the KUGs
in Coma cluster field (see figure~2 of his paper).
The KUGs exist even in the central region of Coma cluster,
though the number is small.
Caldwell \etal (1993) presented a number of spectra
of Coma cluster members with early-type morphologies
to investigate `abnormal' spectra,
which have strong Balmer absorption lines
(see figure~17 of their paper).
These spectra resemble the E+A spectra for galaxies at distant clusters
which are interpreted as post-starburst galaxies
related to the Butcher-Oemler effect
(Dressler, Gunn 1983; Butcher, Oemler 1984).
Among about 500 galaxies of Takase (1980),
which contains both KUGs and non-KUGs down to $m_{pg}$ = 18 in Coma,
we noticed twelve galaxies having the abnormal spectra
or strong \HD\ absorption by inspecting the data given by
Caldwell \etal (1993).
Table~7 tabulated the data of the twelve galaxies;
the galaxy ID-number given in Goldwin, Metcalfe, Peach (1983),
$b$ magnitude, $b-r$ color, and equivalent width of the \HD\ in
the unit of \AA\ given in Caldwell \etal (1993),
and cross-identification with KUG and Markarian catalogs.
Only one object,
KUG~1256+278B,
is included in our total sample of 4050
and its $T$ index is $-$3.0.
$T$ indices of other four KUGs are unknown.
However,
all of them are recognized as early-type galaxies
by Caldwell \etal (1993).

Table~7 shows a clear relation among KUG, Markarian, and post-starburst
galaxies.
The KUG survey picked up all galaxies with abnormal spectra
with EW(\HD\ absorption) $\geq$ 3.5 \AA,
and no galaxies with EW(\HD) $<$ 3.5 \AA\ were recognized as KUGs.
Markarian galaxies correspond to KUGs with the UV degree of M
and have more strict criterion than KUG
in a sense that the objects with EW(\HD) $\geq$ 4.5 \AA\
match the Markarian galaxies.
The three galaxies in table~7 following top two Markarian galaxies
are enough bright to be picked up as the Markarian galaxies.
This means that the KUG catalog
is not a simple extension
to fainter magnitude of the Markarian catalog
and that the KUG survey picked up the post-starburst galaxies
more effectively.
The spectral feature of the post-starburst galaxies,
blue SED (spectral energy distribution) without
strong emission lines,
passes the KUG criteria more effectively.

In the Local Universe,
only several post-starburst galaxies are reported so far
(e.g., Walker, Lebofsky, Rieke 1988; Zabludoff \etal 1996).
The post-starburst galaxies may be common among the blue galaxies
in the Local Universe
as is at intermediate redshifts (e.g., Dressler, Gunn 1988).

% Section 5
\section{Summary}

We analyze 1240 KUGs comparing with 2804 non-KUGs
to investigate the characteristics of the KUGs.

1.~
We introduced the quantitative expressions for the KUG criteria
for the first time.
The boundary colors separating the KUGs from the non-KUGs
are 0.10 and 0.74 in \UBTT\ and \BVTT,
respectively.
The overlapping fractions in colors
between the KUGs and the non-KUGs are 21 to 23\%.
The mean colors in \BVTT\ for objects with the UV degree of
H, M, and L are
0.48, 0.54, and 0.67,
respectively.

2.~
The colors of the KUGs correspond to those of Sc or later-type galaxies
in the Hubble sequence.
The boundary color corresponds to that of a G0 star of the main sequence.

By our quantitative analysis,
the KUGs are clearly shown to be a sample of blue galaxy population.
The results and discussion given for the KUGs,
summarized below,
are for the blue galaxy population in general.

3.~
More than a half of the KUGs with known Hubble-type morphology
are Sb to Scd ($3 \leq T < 7$).
The dwarf irregulars do not dominate in number
because of their low luminosities.

4.~
The KUGs prefer late-type spirals in the Hubble system
and the KUG fraction changes linearly along the Hubble sequence;
it is less than 10\% for E/S0 and more than a half for Sd/Sm.

5.~
In terms of the stellar population
the late-type KUGs are normal galaxies and
the early-type KUGs are peculiar galaxies.
The color difference between the KUGs and the non-KUGs is
negligible for the KUGs with $T$ $\geq$ 5,
while it is significant for the KUGs with $T$ $<$ 5;
the early-type KUGs have much young star populations for their morphologies.
We analyzed the data of \fFIRfB,
a measure of the present star formation activity in galaxies.
It was found that for Sb or earlier-type categories,
the KUG sample shows several times more intense activity
than the non-KUG sample.
In the late-types,
the KUGs have similar star formation history and present
star formation activity to the non-KUGs.
We calculated \LFIRMT\ and found that
the typical value of the present-to-past star formation rate
for the KUGs is 0.4
and nearly constant in any Hubble morphology.

6.~
The early-type KUG sample also contains the post-starburst galaxies
with EW(\HD\ absorption) $\geq$ 3.5 \AA;
such objects are more effectively picked up in the KUG survey
than in the Markarian survey.
It has been argued that
the post-starburst galaxies among the blue galaxies
are found preferably at a cosmological distance.
The KUGs,
the local counterpart of the blue galaxies,
surely contains the post-starburst galaxies,
though the quantitative fraction is not clear so far.

7.~
In the late-type morphology,
the KUGs show more knotty morphology in blue light
than the non-KUGs,
though the stellar populations are similar to each other.
The early-type KUGs have blue knots in their red disks or bulges
and this peculiar morphology is related to the intense
star formation activity for their Hubble type.

8.~
The KUGs are biased to less luminous galaxies.
We analyzed the luminosity functions of the KUG and the non-KUG samples.
For log~\LB\ $<$ 10,
the number density of the KUG and the non-KUG are similar to each other.
For log~\LB\ $\geq$ 10,
the number density of the non-KUG is several times larger than
that of the KUG.
This is also consistent with the selection effect in the KUG fraction
due to distance.
Being less luminous tends to lead to
the intense star formation in earlier Hubble types.
At around the knee of the luminosity function of the KUGs,
most of the KUGs are spiral galaxies.
Dwarf population dominates in fainter class of log \LB\ $\leq$~9.3.

9.~
The fraction of the blue population in a survey
depends on the depth of the survey.
If the survey is enough deep to pick up
most of the dwarf population,
the fraction would be high;
this is another bias to be considered for the fraction of the blue population
in a cosmologically deep survey as well as the evolutionary effect.

\acknowledgments

One of the authors (TTT) acknowledges the Research Fellowships
of the Japan Society for the Promotion of Science for Young Scientists.
This research has also made use of the NASA/IPAC Extragalactic Database (NED)
which is operated by the Jet Propulsion Laboratory, Caltech,
under contact with the National Aeronautics and Space Administration.
Finnaly,
We are grateful to the anonymous referee for improving the paper.

\clearpage

\figcaption[fig1.ps]
{
Histograms of colors for the KUG and the non-KUG samples
in \UBVTT\ system.
Solid one indicates the KUG sample and the dashed one
indicates the non-KUG sample.
Each histogram is normalized to the total number of
the each sample.
The abscissa is for the color and the color system
used is shown in the upper left corner in the histogram.
The ordinate is for the relative frequency in the unit of \%.
An arrow indicates the boundary color separating the KUG
from the non-KUG samples.
The characteristics of the histograms are also
tabulated in table~1.
\label{fig1}}

\figcaption[fig2a.ps, fig2b.ps, fig2c.ps, fig2d.ps]
{
(a)
\UBVTT\ color-color diagram for the KUG and the non-KUG samples.
Circles indicate the KUG sample and
crosses indicate the non-KUG sample.
The abscissa indicates \BVTT\ and
the ordinate indicates \UBTT\ colors.
Dashed line indicates the color boundary
separating the KUG from the non-KUG samples,
\UVTT\ = 0.83,
given in table~1.
(b)
Color variation along the Hubble morphological sequence
shown in \UBVTT\ plane.
The color data for each Hubble type was derived
from a combined sample of the KUGs and non-KUGs.
The data is also tabulated in table~2.
The dashed line, the same as in figure~2a,
shows the boundary color.
The colors of the KUGs correspond to those
of Sc or later-type galaxies.
(c)
The same as figure~2b,
but in \UBVTo\ plane.
Dashed line indicates the boundary color
of \UVTo\ = 0.64.
Colors of the main sequence stars are plotted.
The color boundary corresponds to that of G0 stars,
though originally the color of the KUG criterion
was intended to be A0 stars.
\label{fig2}}

\figcaption[fig3a.ps, fig3b.ps]
{
Color variation along the UV degree, H, M, and L.
(a) Histogram shown in \UBTT\ and (b) in \BVTT.
Each histogram is normalized to the total number of each sample.
The ordinate indicates the relative frequency in \%.
There is a correlation between the UV degree and
the total colors,
though the separation of color distribution is not clear.
The mean color and the standard deviation of the distribution
in each color system is tabulated in table~3.
Note that the color with the UV degree of L
is bluer than the boundary color
separating the KUG from the non-KUG samples.
\label{fig3}}

\figcaption[fig4.ps]
{
The fraction of the KUG in a bin of the $T$ index
along the Hubble sequence
drawn using the data in table~4.
Solid histogram shows the number of the KUG
and dashed one shows the number of the total samples.
The scale of the number is shown in left-side ordinate.
The abscissa indicates the Hubble morphological type,
$T$ index.
A solid broken line shows the KUG fraction
in a bin of the $T$ index.
The scale of the fraction in \% is shown
in right-side ordinate.
The fraction is monotonously increasing along the Hubble type.
\label{fig4}}

\figcaption[fig5.ps]
{
Variation of the relative frequency of the UV degree
in each $T$ index along the Hubble sequence.
The fraction of the higher UV degrees is monotonously
increasing with $T$ index.
The data is also tabulated in the last three columns of table~4.
\label{fig5}}

\figcaption[fig6.ps]
{
\BVTT\ colors of objects with the UV degree of M
for different Hubble morphological types.
The ordinate indicates the relative frequency in each histogram.
The most lower panel shows the histogram for the total sample
with the UV degree M.
The mean color and the standard deviation of the distribution
of each histogram is tabulated in table~5.
\label{fig6}}

\figcaption[fig7.ps]
{
Radial velocity \VGSR\ distribution of a combined sample
of the KUGs and the non-KUGs.
The velocity data was taken from the RC3.
The number of data used is 846.
The solid curve indicates the canonical $N$-$z$ relation
which is expected for a uniform distribution of galaxies.
\label{fig7}}

\figcaption[fig8.ps]
{
\BTo\ distribution of a combined sample of the KUGs and the non-KUGs.
The data was taken from RC3.
The number of data used is 3391.
A single power-law of log~$N \propto$ (0.5\BTo)
follows down to \BTo\ = 14.2.
\label{fig8}}

\figcaption[fig9.ps]
{
Log~$L_{B}$ luminosity functions for
the KUG and the non-KUG samples.
The solid histogram is for the KUGs and the dashed one
is for the non-KUGs.
The solid curve indicates the Schechter-type canonical luminosity function.
\label{fig9}}

\figcaption[fig10a.ps, fig10b.ps]
{
The luminosity functions of the KUGs for
four morphological subsamples;
(a) samples with $-6 \leq T < 2$ (Sa or earlier) and
$2 \leq T < 5$ (Sb or Sbc), and
(b) $5 \leq T < 7$ (Sc or Scd) and
$7 \leq T \leq 11$ (Sd or later).
The dashed histogram shows the luminosity funciton
of the total KUG sample
as shown in figure~9.
\label{fig10}}

\figcaption[fig11.ps]
{
The variation of the relative fraction of the KUGs
in the four morphological categories along \LB;
sample with $-6 \leq T < 2$ (denoted as E -- Sa in the panel),
$2 \leq T < 5$ (Sb),
$5 \leq T < 7$ (Sc),
and $7 \leq T \leq 11$ (Sd -- Im).
\label{fig11}}

\figcaption[fig12.ps]
{
The variation of the KUG fraction as a function of
\VGSR\ in a bin of 1000~km~s$^{-1}$.
0 --~1000~km~s$^{-1}$ bin contains galaxies
with \VGSR\ $<$~1000 [km~s$^{-1}$],
1000 --~2000~km~s$^{-1}$ bin contains galaxies
with 1000 $\leq$~\VGSR\ $<$~2000 [km~s$^{-1}$] and so on.
\label{fig12}}

\figcaption[fig13.ps]
{
Distribution of far-infrared-to-$B$ band flux ratio
for each Hubble morphological type;
E/S0: $T < 0$,
Sa: $0 \leq T < 2$,
Sb: $2 \leq T < 4$,
Sc: $4 \leq T < 6$,
Sd: $6 \leq T < 8$,
and Sm/Im: $8 \leq T \leq 11$.
The left-side panel indicates the KUG sample
and the right-side panel indicates the non-KUG sample.
The abscissa indicates the flux ratio in logarithmic scale
and the ordinate indicates the relative frequency within each panel
in \%.
The solid histogram shows the sample with measured values of \fFIRfB,
and the dashed one shows the sample with only upper-limit values
of \fFIRfB.
Note that these histograms are simple frequency of our sample.
\label{fig13}}

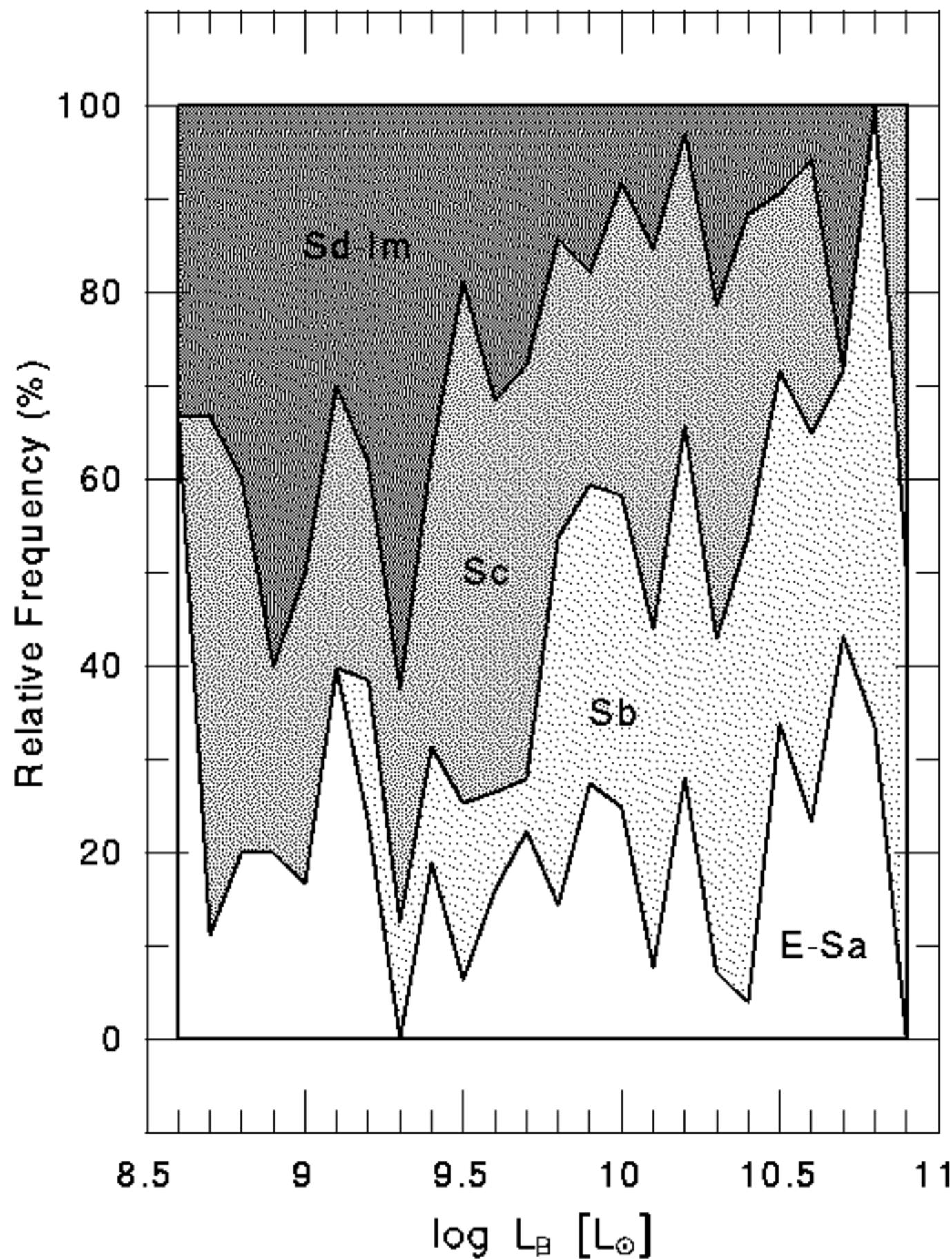
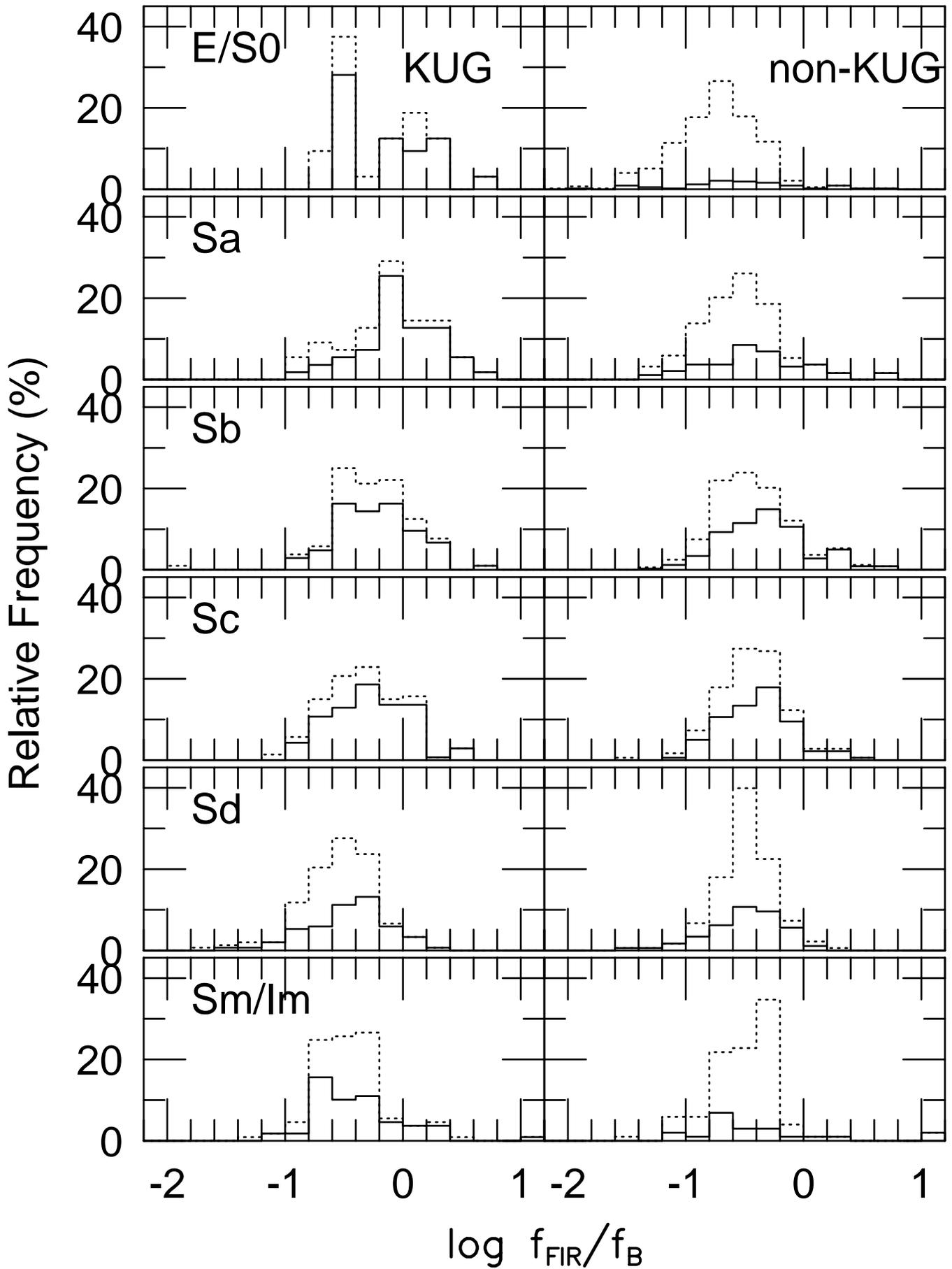
\end{document}